\documentclass{article}

\usepackage{longtable}
\usepackage{url}
\usepackage{graphicx}

\title{Blueprint and Evaluation Instruments for a Course on Software Engineering for Sustainability}

\author{Birgit Penzenstadler\\
CSU Long Beach\\
Long Beach\\
USA \\
\url{birgit.penzenstadler@csulb.edu}
\and
Stefanie Betz\\
Karlsruhe Insitute of Technology\\
Karlsruhe\\
Germany\\
\url{stefanie.betz@kit.edu}
\and
Colin C. Venters\\
University of Huddersfield\\
Huddersfield\\
UK\\
\url{c.venters@hud.ac.uk}
\and
Ruzanna Chitchyan\\
University of Bristol\\
Bristol\\
UK\\
\url{r.chitchyan@bristol.ac.uk}
\and
Jari Porras\\
LUT\\
Lappeenranta\\
Finland\\
\url{jari.porras@lut.fi}
\and
Norbert Seyff\\
FHNW\\
Windisch\\
Switzerland\\
\url{norbert.seyff@fhnw.ch}
\and
Leticia Duboc\\
La Salle\\
Barcelona\\
Spain\\
\url{lduboc@salleurl.edu}
\and
Christoph Becker\\
University of Toronto\\
Toronto\\
Canada\\
\url{christoph.becker@utoronto.ca}
}

\begin{document}
\maketitle

\newpage
\begin{abstract}
We report on a summer school course on Software Engineering for Sustainability (SE4S). We provide a detailed blueprint of the contents taught and its evaluation with the instruments that were used.
\end{abstract}

\section{Intro and Context}

Sustainability has become an important concern across many disciplines, and software systems play an increasingly central role in addressing it. However, teaching students from software engineering and related disciplines to effectively act in this space requires interdisciplinary courses that combines the concept of sustainability with software engineering practice and principles. Yet, presently little guidance exist on which subjects and materials to cover in such courses and how, combined with a lack of reusable learning objects. 

Penzenstadler et al.~\cite{penzenstadler2018icse} describe a summer school course on Software Engineering for Sustainability (SE4S). In the paper at hand, we provide a detailed blueprint for this course, in the hope that it can help the community develop a shared  approach and methods to teaching SE4S. The course blueprint, availability of used materials and report of the study results make this course viable for replication and further improvement.

Furthermore, the paper at hand presents in detail the instruments that were used for the evaluation of the course. We used three evaluation instruments: 
\begin{enumerate}
\item a pre-survey, 
\item a post-survey, and 
\item a list of evaluation criteria for the analysis of the artifacts produced by the students in the course.
\end{enumerate}

\section{The SE4S Course}

The week-long course is designed as part of a summer school held at the Lappeenranta University of Technology (LUT) in Finland.

The learning objectives and their assessment are as follows:

\begin{enumerate}
\item 
\textbf{Sustainability concepts and principles}: Develop an understanding of the concept of sustainability and its different dimensions and orders of effect and an ability to transfer these concepts to other application domains.
\textit{Assessment}: Students will demonstrate their mastery of sustainability concepts in demonstrating the transfer to a different application domain in their team project documentation.
\item 
\textbf{Requirements Engineering}: Students develop an understanding of the basics of requirements engineering, they understand and are able to apply stakeholder modeling, goal modeling, process modeling, use case modeling, and SysML.
\textit{Assessment}: Students demonstrate their mastery of requirements engineering by developing a consistent specification that includes stakeholders, goals, process model, use cases, and SysML diagrams.
\item
\textbf{Systems thinking}: An understanding of the mindset of and the general principles of systems thinking, including holistic viewpoints and iterative development. Understand and be able to reason about long-term effects that a system under development may have on the environment and on society.
\textit{Assessment}: Students demonstrate their knowledge in taking a bigger picture perspective and holistic viewpoint in their rich picture. Demonstrate the reasoning in providing an analysis to that regard for the system under development and pointing out risks that may or may likely occur in the future given certain conditions.
\item
\textbf{Design thinking}: Understand and be able to apply design thinking on (complex) problems, which requires alternating between narrowing down and opening up the perspective.
\textit{Assessment}: Demonstrate the application of design thinking on a local problem to demonstrate understanding and transfer of the concepts of iterative development in innovation in their project.
\end{enumerate}

The subject areas and modules are detailed in Tab.~\ref{tab:modules}. For each, we provide content, example key references, and rationale for their inclusion in the course. 
In the following, we add more details on the content of each lecture module:

\begin{enumerate}
\item \textbf{Sustainability Foundations}: This module explains the fundamental concepts for talking about, analyzing, and designing for sustainability. This includes the topics of Dimensions of sustainability, orders of effect, application domains, and the Sustainable Development Goals. Recommended readings are~\cite{hilty2015ict,becker2015,griggs2013policy}. The module provides a common basis for scoping sustainability within this course and puts the concept into a larger perspective.
\item \textbf{Principles of sustainability design}: This module delivers and explains the principles of substitution, decoupling, and dematerialization (see Hilty and Aebischer~\cite{hilty2015ict}) as well as 
Software Engineering for Sustainability examples~\cite{penzenstadler2014infusing,chitchyan2015evidencing}. 
\item \textbf{Rich pictures}: This module introduces the technique of rich pictures, a method for scoping and high-level domain modeling, as proposed by Monk and Howard~\cite{monk1998methods}. Rich pictures are a simple, non-technical method to illustrate the vision for a system or scoping of a problem in its surrounding application domain and operational context. The connection to sustainability lies in the fact that rich pictures provide a more coarse-grained picture of the larger context in which a system resides, and makes the relationships between stakeholders and problem domain clear.
\item \textbf{Stakeholders and goal models}: The stakeholders that for the first time show up in the rich picture are now analyzed in more detail. Based on the reference model of Penzenstadler et al.~\cite{penzenstadler2013advocate}, the stakeholder model makes sure all relevant roles are included. The goal modeling is based on the generic sustainability goal reference model by Penzenstadler and Femmer~\cite{penzenstadler2013generic} and provides a basis for consensus, conflict identification and trade-offs with specific focus on the different aspects of sustainability. The module is an introduction to concepts, roles, reference models, analysis, and notations for both of the models.
\item \textbf{Process modeling}: This modules gives an introduction to concepts of and notation for business process modeling~\cite{ould1995business,larschBDMB16} and thereby provides the transition from high-level goals to the operationalization of these objectives in executable processes. This is the important tie-in of sustainability goals to the implementation - the step needs to be made traceable for later assessment.
\item \textbf{Software engineering}: We provide an overview of the general Software Engineering process phases and introduction to use cases~\cite{kotonya1998requirements}, such that usage behavior can now be tied in with sustainable human computer interface design. This complements the technology-agnostic processes with the technology-aware perspective by describing the interaction between user and system.
\item \textbf{SysML}: The introduction to SysML provides the foundation for analyzing and designing complex systems according to~\cite{friedenthal2008omg}. The Systems Modeling Language is a widely used general purpose language for modeling and verifying software systems and its widespread usage across industry makes it a good exemplar for this part of the course.
\item \textbf{Design Thinking}: The d.school's~\cite{d.school} highly interactive workshop on design thinking facilitates transition into rapid prototyping. It's a hands-on crash course tutorial in design thinking that has students team up and work on physical prototypes. User engaging with physical prototypes has brought about many sustainability concerns in human computer interaction design and proved useful to get literally ``a better feeling'' for the products and services under development.
\item \textbf{Sustainability Analysis}: The overview of all impacts of a system from short-term to long-term enables software engineers to take a wider perspective and better judge the consequences of their choices during development. We introduce the analysis and estimation of a system's impact according to the sustainability dimensions and order of effect as described in~\cite{becker2016}. This simple tool with a one-page template enables students to perform a high-level long-term assessment of a system under development within half an hour. While the assessment outcomes of such a short time should not be used as decision basis, they point to the most important questions and potential consequences of the long-term use of a system that should be investigated.
\end{enumerate}

\begin{table}[htbp]
\centering \footnotesize
\caption{Subject areas and modules}\label{tab:modules}
\begin{tabular}{|p{2cm}|p{4cm}|p{1cm}|p{5cm}|}\hline
\textbf{\textit{Module}} & \textbf{\textit{Content}} & \textbf{\textit{Key refs.}} & \textbf{\textit{Rationale for inclusion}}\\\hline
Sustainability foundations
& Dimensions of sustainability, orders of effect, application domains, Sustainable Development Goals
& \cite{hilty2015ict,becker2015,griggs2013policy} 
& Provides a common basis for scoping sustainability within this course and puts the concept into a larger perspective.\\\hline
Principles of sustainability design
& Principles of substitution, decoupling, and dematerialization;
Software Engineering for Sustainability examples
& \cite{hilty2015ict,penzenstadler2014infusing,chitchyan2015evidencing} 
& Introduces the principles that can be used for thinking of system ideas for the projects to be developed during the course.\\\hline
Rich pictures
& Rich picture method for scoping and high-level domain modeling
& \cite{monk1998methods} 
& Rich pictures are a simple, non-technical method to illustrate the vision for a system or scoping of a problem in its surrounding application domain and operational context. \\\hline
Stakeholder and goal models
& Introduction to concepts, roles, reference models, analysis, and notations for both of the models.
& \cite{penzenstadler2013advocate,penzenstadler2013generic}
& Forms the basis for eliciting requirements for a chosen project idea. The stakeholder model makes sure all relevant roles are included, the goal model provides a basis for consensus, conflict identification and trade-offs.\\\hline
Process \newline modeling
& Introduction to concepts of and notation for business process modeling
& \cite{ould1995business,larschBDMB16} 
& Provides the transition from high-level goals to the operationalization of these objectives in executable processes.\\\hline
Software \newline Engineering 
& Overview of the general Software Engineering process phases and introduction to use cases
& \cite{kotonya1998requirements} 
& Complements the technology-agnostic processes with the technology-aware perspective by describing the interaction between user and system.\\\hline
SysML
& Introduction to SysML for analyzing and designing complex systems
& \cite{friedenthal2008omg}
& The Systems Modeling Language is a widely used general purpose language for modeling and verifying software systems.\\\hline
Design \newline Thinking 
& Hands-on crash course tutorial in design thinking
& \cite{d.school}
& The d.school's highly interactive workshop on design thinking facilitates transition into rapid prototyping.\\\hline
Sustainability analysis
& Introduction to the analysis and estimation of a system's impact according to the sustainability dimensions and order of effect
& \cite{becker2016}
& The overview of all impacts of a system from short-term to long-term enables software engineers to take a wider perspective and better judge the consequences of their choices during development.\\\hline
\end{tabular}
\end{table}

\begin{figure}[htbp]
\centering
\includegraphics[width=\textwidth]{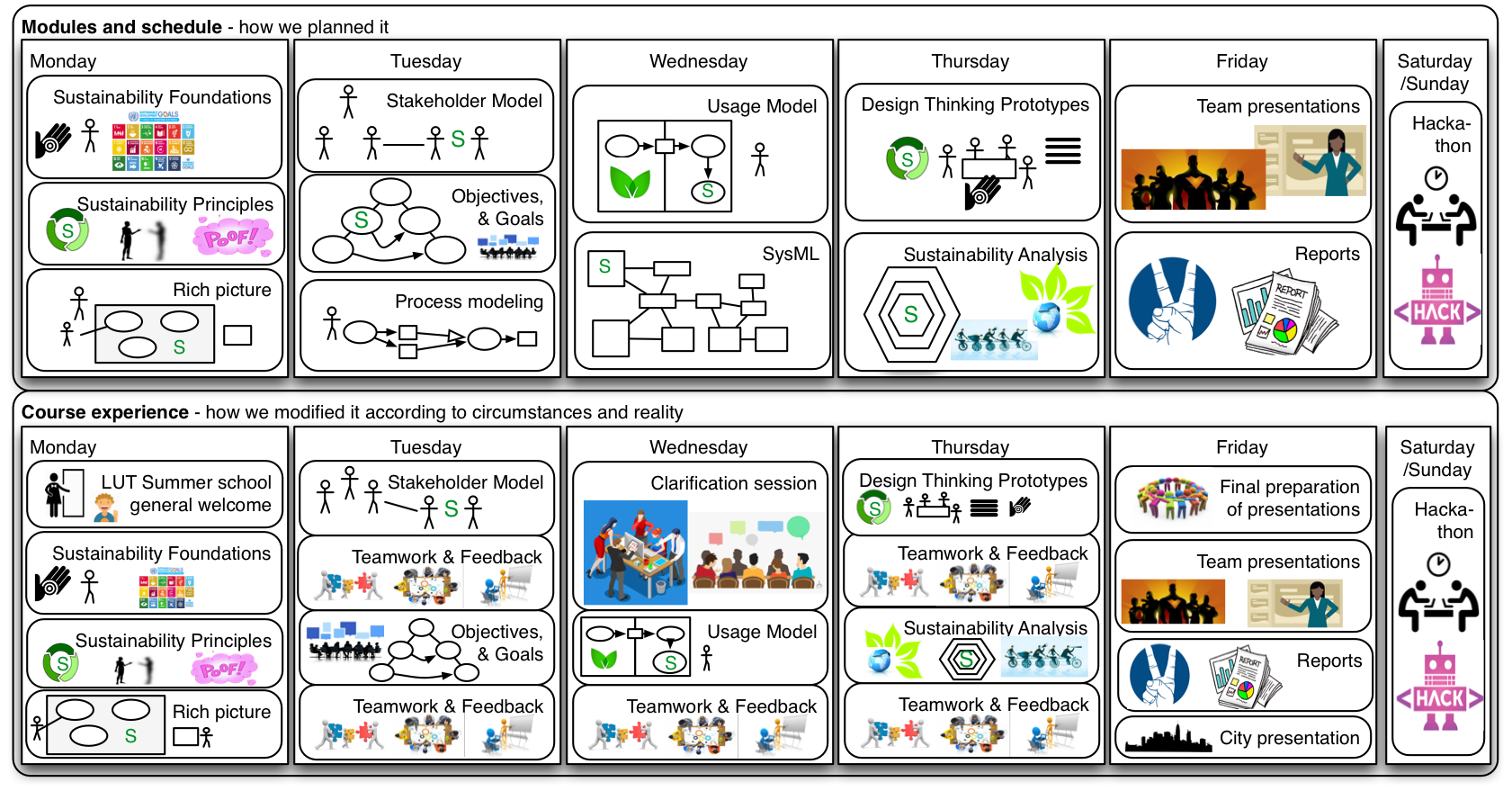}
\caption{SE4S course planned schedule and actual experience}\label{fig:schedule}
\end{figure}

The planned schedule of modules of the course and the how we modified it according to circumstances are depicted in Fig.~\ref{fig:schedule}.

\clearpage

\section{Evaluation Instruments}

We used a pre- and post-survey and a report for the students' self-assessment and learning perceptions, and for the external assessment in form of artifact analysis we used a criteria catalogue.

\subsection{Pre-Survey before the SE4S course}

A short pre-survey (Table~\ref{tab:survey}) evaluates whether our (informal and internal) hypothesis about the characteristics of the student population would hold, and to be able to better tailor the course to the student population that decided to register for the course in that instance. 

The survey contained questions on their familiarity and experience with the general software engineering process, requirements elicitation and modeling, UML diagrams, SysML, IFML (Interaction Flow Modelling Language), Attribute-Driven Design (ADD), user interface development, rapid prototyping, design thinking, systems thinking, and computational thinking.

\begin{table}[h!]
\centering \footnotesize
\caption{Pre-Survey}\label{tab:survey}
\begin{tabular}{|p{\textwidth}|}
\hline
\begin{enumerate}\itemsep0pt
\item Do you know the general software engineering process (the phases, roles)?
\item Have you taken a course on software engineering?
\item Have you ever developed a software project (by yourself or in a team)? Please briefly describe what you did and which technologies you used (bullet points are fine).
\item Have you written down requirements, e.g. in natural language?
\item Have you ever modeled requirements, e.g. in use cases?
\item Have you used UML diagrams for software design?
\item Have you used sysML for systems design?
\item Have you used IFML (Interaction Flow Modelling Language) for software design?
\item Have you used Attribute-Driven Design (ADD) for software architecture design?
\item Have you ever developed a user interface / graphical interface? 
\item Have you used rapid prototyping?
\item Have you heard of ``design thinking''? If so, have you ever used it?
\item Have you heard of systems thinking? If so, have you ever applied it?
\item Have you heard of computational thinking? 
\item Have you already signed up for the hackathon on the weekend August 5/6?
\end{enumerate}
\\\hline
\end{tabular}
\end{table}

\clearpage

\subsection{Post-Survey after the SE4S course}

We used a survey (Table~\ref{tab:postsurvey}) that consisted of several sections, one part dedicated to the SE4S course, one part on ethics perceptions, and one part on value ratings.\footnote{Penzenstadler et al.~\cite{penzenstadler2018icse} analyzes the parts of the survey dedicated to the SE4S course due to space limitations.} 
The part specific for the SE4S course was composed by a background section, the motivation for the course selection, a section on their notions of sustainability, software and software engineering, their perceptions of the course content, and reflections on the course.

\begin{longtable}{|p{\textwidth}|}\caption{Post-Survey}\label{tab:postsurvey}\\\hline
\textbf{Background} \\
2. Gender\\
3.	What is your age range?\\
4.	What is your highest education level?\\
5.	What is your highest qualification?\\
6.	What is your discipline?\\\hline
\textbf{Motivation for Course Selection}\\
7.	Why did you choose this course?\\
8.	What did you expect to learn from this course?\\\hline
\textbf{Notions of Sustainability, Software and Software Engineering}\\
9.	What did you think sustainability is before you started this course?\\
10.	What do you think sustainability is now, after completing the course? \\
11.	How much has your perception of sustainability changed as result of this course? \\
12.	Before starting this course, did you think software could help with sustainability? If yes, then how?\\
13.	After completing this course, how has your perception changed: do you think software can help with sustainability? And If yes: how?\\
14.	Before starting this course how much did you think can software help with achieving sustainability?\\
15.	After completing this course, how much, do you think, software can help with achieving sustainability?\\
16. What element of the course changed your perception, if any?\\
17.	What did you think software engineering was before you started this course?\\
18.	What do you think software engineering is now, after completing the course?\\
19.	Before starting this course, how did you think software \textit{engineering} could help with sustainability?\\
20.	After completing this course, how do you think software \textit{engineering} can help with sustainability?\\
21.	Before starting this course how much, did you think, can software \textit{engineering} help with achieving sustainability?\\
22.	After completing this course, how much, do you think, software \textit{engineering} can help with achieving sustainability?\\
23.	What did you think ``software engineering for sustainability'' was before completing this course?\\
24.	What do you think ``software engineering for sustainability'' is now, after completing the course?\\
25.	What helped you most to link the notions of software engineering and sustainability? (You can consider the material taught in this module as well as any other sources).\\\hline 
\textbf{On Course Content}\\
26.	What three key things related to sustainability did you learn in this course?\\
27.	What three key things related to software engineering did you learn from this course?\\
28.	How useful do you think the following techniques are for Software Engineering for Sustainability? (Rating rich picture, stakeholder modeling, goal modeling, use case modeling, design thinking and sustainability analysis on a Likert scale (1: not at all - 5: all the time))\\
29.	Which of the following techniques would you apply in your further work/research? (rating of same techniques on same Likert scale)\\\hline
\textbf{Reflections}\\
30.	What would you like to do next with the knowledge you learned in this course?\\
31.	What would you like to learn next about sustainability and/or software engineering?\\
32.	What did you especially like about this course?\\
33.	What did you especially dislike about this course?\\\hline
\end{longtable}

\clearpage

\subsection{Criteria for the analysis of the artefacts}\label{sec:criteria}

\begin{figure}[h!]
\includegraphics[width=\columnwidth]{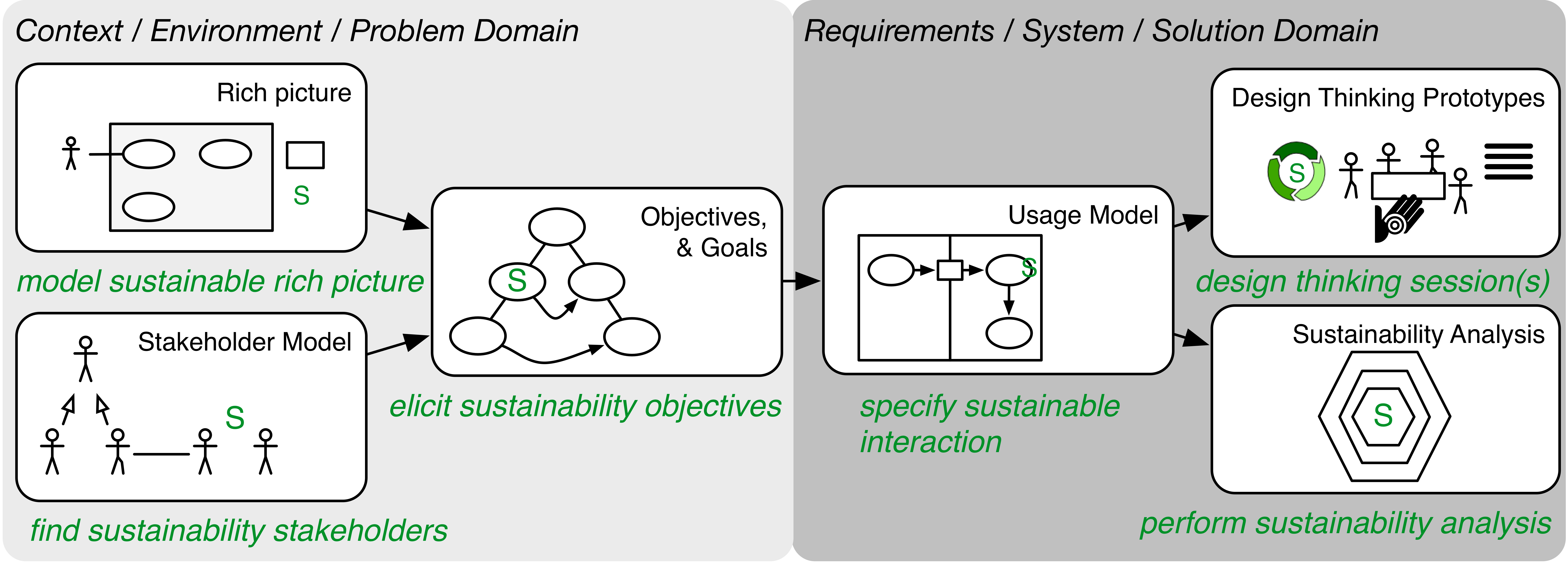}
\caption{RE4S artefact model used in the course}\label{fig:re4s}
\end{figure}

Over the course of the week, the objective was to develop a specification according to a small requirements artifact model as well as some prototypes or mock-ups. An overview of the artifact model to be produced is given in Fig.~\ref{fig:re4s}. The artifacts are a rich picture, a stakeholder model, a goal model, a use case overview model, design thinking prototypes, and a sustainability analysis diagram.
For the artifact analysis, we used a list of jointly elaborated quality criteria to structure their analysis (Table~\ref{tab:criteria}). For each artifact, there is a number of questions and criteria to be considered by the evaluators.

\begin{longtable}{|p{\textwidth}|}\caption{Criteria for the analysis of the artefacts}\label{tab:criteria}
\\\hline
\textbf{Overall} \\
\begin{itemize}
\item Does the project address the chosen sustainability challenge well?
\end{itemize}
\\\hline 
\textbf{Stakeholder model}\\
\begin{itemize}
\item Have all major stakeholder groups been considered?
\item Is the diagram easy to understand?
\item Are there well-organized hierarchies?
\item Are the relationships between the stakeholders clear?
\item Have stakeholders been analyzed and described (to an adequate degree of detail)?
\end{itemize}
\\\hline \pagebreak
\textbf{Goal model}\\
\begin{itemize}
\item Is the goal model well structured?
\item Are they broken down and refined into sub-goals where possible?
\item Are the influences and conflicts between sustainability and other goals represented?
\item Are all stakeholders? main concerns represented by goals?
\item Is a good accompanying explanation provided for the diagram?
\end{itemize}
\\\hline
\textbf{Rich picture}\\
\begin{itemize}
\item Can I (picturing myself as a stakeholder) easily understand what the problem is about?
\item Is the vision described and illustrated to an adequate degree of detail?
\end{itemize}
\\\hline
\textbf{Usage model}\\
\begin{itemize}
\item Is a use case overview diagram provided and includes all important use cases? 
\item Does it have a system boundary, an actor outside that boundary, and relations from the actor to all use cases?
\end{itemize}
\\\hline
\textbf{Sustainability analysis}\\
\begin{itemize}
\item Are all dimensions analyzed?
\item Are all orders of effect analyzed?
\item Are all orders of effect analyzed for all dimensions?
\item Are cross-relations denoted between the effects? Does every second order effect have a first order effect it expands on? Does every third order effect have a second order effect it expands on?
\item Do the predictions in the individual fields describe actually possible effects of the system on its environment (in the opinion of the reviewer)?
\end{itemize}
\\\hline
\end{longtable}

\clearpage

\section{Conclusion}

This paper provides a detailed blueprint for a course on software engineering for sustainability.
The authors are planning to replicate this course every year in different locations and report on the continued evaluation.
We hope the blueprint and evaluation instruments serve further researchers and educators in developing their own courses on software engineering for sustainability.

\bibliographystyle{plain}
\bibliography{bib.bib}
\end{document}